\documentclass[]{hdsr}
\usepackage{csquotes}
\usepackage{todonotes}
\usepackage{soul}

\graphicspath{{figs3/}}

\begin{document}

\newgeometry{bottom=1.5in}

\volumeheader{0}{0}{00.000}

\definecolor{lightblue}{rgb}{.80,.9,1}
\sethlcolor{lightblue}
\renewcommand\hl[1]{#1}

\begin{center}

  \title{Demonstrations of the Potential of AI-based Political Issue Polling}
  \maketitle

  \thispagestyle{empty}
  
  \vspace*{.2in}

  \begin{tabular}{cc}
    Nathan E. Sanders\upstairs{\affilone,*}, Alex Ulinich\upstairs{\affiltwo}, Bruce Schneier\upstairs{\affilthree}
   \\[0.25ex]
   {\small \upstairs{\affilone} Berkman Klein Center, Harvard University, 23 Everett St \#2, Cambridge, MA 02138} \\
   {\small \upstairs{\affiltwo} Mountain View High School, 3535 Truman Avenue, Mountain View, CA 94040} \\
   {\small \upstairs{\affilthree} Harvard Kennedy School, 79 JFK Street, Cambridge, MA USA 02138 }
  \end{tabular}
  
  \emails{
    \upstairs{*}nsanders@cyber.harvard.edu
    }
  \vspace*{0.4in}

\begin{abstract}
Political polling is a multi-billion dollar industry with outsized influence on the societal trajectory of the United States and nations around the world.
However, in recent years it has been severely challenged by rising nonresponse rates and other factors that stress its cost, availability, and accuracy.
At the same time, artificial intelligence (AI) chatbots such as ChatGPT have become highly compelling stand-ins for a wide range of human behavior, powered by increasingly sophisticated large language models (LLMs).
Because these LLMs are trained on huge corpora of writing by diverse people captured from across the Internet, they are potentially capable of representing a wide range of beliefs on many policy issues.
Could AI chatbots be an effective tool for anticipating public opinion on controversial issues to the extent that they could be used by campaigns, interest groups, and polling firms?

We have developed a prompt engineering methodology for eliciting human-like survey responses from ChatGPT, which simulate the response to a policy question of a person described by a set of demographic \hl{and ideological} factors, and produce both an ordinal numeric response score and a textual justification.
We execute large scale experiments using this method, querying GPT for thousands of simulated responses at a cost more than three orders of magnitude lower than human surveys.
We compare this simulated data to human issue polling data from the Cooperative Election Study (CES).

We find that ChatGPT is effective at anticipating both the mean level and distribution of public opinion on a variety of policy issues such as abortion bans and approval of the US Supreme Court, particularly in their breakdown along \hl{ideological} lines (correlation typically $>85\%$).
However, it is much less successful at anticipating demographic (age, race, and gender) differences between respondents.
Moreover, ChatGPT tends to overgeneralize its conception of ideological differences to new policy issues that arose after its training data was collected, such as American support for involvement in the war in Ukraine.
Our work has implications for our understanding of the strengths and limitations of the current generation of AI chatbots as virtual publics or online listening platforms, future directions for LLM development, and applications of AI tools to the political domain.
\end{abstract}
\end{center}

\vspace*{0.15in}
\hspace{10pt}

\copyrightnotice

\section{Introduction}
\label{sec:intro}

While survey experiments and polling have been powerful tools for political campaigns, parties, and advocacy organizations in the US and around the world for centuries \citep{splichal2022data}, in recent years the cost and difficulty of operating polls has grown dramatically.
Political polling firms commonly recruit panels intended to be representative of, and to achieve high coverage of, their targeted population, such as eligible voters nationally or likely voters in a voting district. 
Reaching these populations has become harder primarily because of the growth in survey nonresponse internationally: the failure to contact or refusal of potential participants to be surveyed due to factors such as lack of time, disinterest, and distrust \citep{LuitenHoxde20}.
Moreover, the migration of respondents to new technologies such as cell phones and the Internet, which have uneven and evolving penetration and usage across regions and demographic groups, has constrained the coverage of survey samples \mbox{\citep{berinsky17}}.

These effects have generated simultaneous challenges for the quality and cost of political polling, as biases in political engagement and hyper-polarization manifest on response rates (\cites{olson20}{cavari2023survey}).
A vast literature has developed on statistical methodologies for designing and postprocessing survey data to overcome these challenges, including methods such as demographic weighting and poststratification (see e.g. \cites{kennedy2018evaluation}{isakov2020towards}{berinsky17}).
In particular, pollsters have explored methodologies that enable meaningful public opinion research from digital platforms such as Facebook and other social media platforms, where traditional techniques of probability sampling cannot be applied because of the lack of a conventional sampling frame and researcher-controlled contact mechanism \mbox{(\cites{murphy2014social}{schneider2022s})}.
These various methodologies seem to have been successful at maintaining the predictive accuracy of election polling thus far, even as nonresponse has proliferated \citep{jennings2018election}, and yet there is widespread interest in finding transformative new models for measuring public opinion that could lead to more cost-effective, sustainable, and more reliable polling results (\cites{forbes_polling_2020}{jacobin_polling_2020}{pew_polling_2023}{hill_polling_2020}{fivethirtyeight_polling_2021}{atlantic_polling_2020}).

As statistical methodologies have come to play a critical role in collecting, processing, and interpreting political polling data, machine learning (ML) and artificial intelligence (AI) systems may further revolutionize this domain.
In particular, large language models (LLMs) such as ChatGPT, which can be incorporated into AI chatbots and other systems capable of providing human-like responses to natural language prompts, have a wide variety of potential applications in democratic processes, such as assisting lobbying firms (\cites{sanders_schneier_2023_nyt}{nay2023large}), helping citizens and stakeholders to formulate and advocate for their opinions \citep{schneier_2023_slate}, facilitating connections between candidates and voters \citep{sanders_schneier_2023_atlantic}, and even helping humans social engineer or hack political systems (\cites{schneier_coming_2021}{sanders2021machine}{sanders_2023_techrev}).
Already, researchers have experimented with a variety of social science research and public polling applications of LLMs, such as coding open-ended survey responses \citep{mellon2022does}, \hl{providing synthetic participants for human subjects research \protect{(\cites{pmlr-v202-aher23a}{dillion2023can})}, mimicking consumer responses to market factors like price sensitivity \protect{\citep{brand2023using}}}, inferring the ideology of a politician \citep{wu2023large}, \hl{representing the personality traits of psychological profiles \protect{\citep{jiang2023personallm}}}, simulating economic behavior \citep{horton2023large}, \hl{simulating feelings towards political parties and groups} \citep{bisbee2023artificially}, and simulating election results \citep{argyle2022out}. 

Because they are trained on wide Internet corpora including opinion writing from a diverse range of people, LLM's have a compelling ability to represent different perspectives and to perform a wide range of tasks without specialized training (\cites{brown2020language}{kojima2022large}{y2022large}).
We therefore hypothesize that they may be effective at generating individualized responses to policy preference questions that can account for the same factors that influence human respondents, such as demographics \hl{and ideology}. 

However, the nature of LLMs limits their potential effectiveness as opinion sampling tools.
Like platforms such as social media, AI chatbots do not have well defined sample frames or well understood coverage characteristics.
Moreover, unlike true survey platforms, using LLMs does not actually involve any solicitation of opinion from an authentic human individual.
Instead, LLMs generate a response predicted to be most acceptable to the user on the basis of a training process such as reinforcement learning with human feedback \mbox{\citep{ziegler2019fine}}, which may therefore reflect the incomplete \hl{or biased} properties of its training dataset.
Some specific biases of Internet corpora-trained LLMs are coming in to focus.
One study attempted to assess the age and gender characteristics of ChatGPT by prompting it to express a demographic profile, finding that its responses are biased towards a young ($<30$ years old) and female profile \mbox{\citep{miotto2022gpt}}.
Other investigators identified that an earlier model, GPT-2, is biased in its representation of the opinions of people from nations underrepresented in Internet usage \mbox{\citep{venkit2023nationality}}.
Regardless of their ability to reflect the perspectives of a given demographic group, AI models may also exhibit bias in the text they generate; for example, in an analysis of the BERT model, researchers found that neural embeddings learn harmful stereotypes about persons with disabilities \mbox{\citep{hutchinson2020social}}.
\hl{For applications to survey research specifically, the ability to reproduce such biases and stereotypes may enable the LLM to authentically reflect the beliefs of the human population, even if inaccurate or morally negative; but only to the extent that its training data realistically reflects these sentiments.}

In this work, we seek to test the capability of current generation AI tools to accurately reflect distributions of public opinion, and to expose insight into its effective sociodemographic coverage as a polling instrument, using a generally available LLM and real public opinion survey questionnaires. 
We have developed experimental methods (\S\ref{sec:meth}) to prompt the AI chatbot ChatGPT to generate public polling-like responses such that it can simulate a survey panel. 
We test the model's ability to reflect the shift in valence between demographic \hl{and ideological} groups across a variety of issues, as well as reasonably reproduce the key arguments appealed to by each \hl{subgroup} (\S\ref{sec:res}). 
We provide an interpretation of this capability in the context of prior Internet-assisted approaches to public opinion research, discuss the limitations of this approach and the current generation of tools, and the implications these capabilities may have as they improve (\S\ref{sec:disc}), before concluding (\S\ref{sec:conc}).

\section{Methods}
\label{sec:meth}

We explore the viability of AI language models to simulate public opinion polling responses by developing a system that automates querying an LLM based on the questionnaire of a survey previously given to people, so that the resulting AI responses are aligned and comparable to human data.\footnote{We will publish the code associated with this work at the time the article is accepted.}

\subsection{Large Language Model}
\label{sec:meth:llm}

We use the OpenAI Chat Completion API endpoint, through OpenAI's \textit{openai} python library,\footnote{\url{https://github.com/openai/openai-python}} to query the \textit{gpt-3.5-turbo-0301} LLM for polling responses. 
This model was the most recent model from OpenAI optimized for chat applications and made generally available as of April 2023\footnote{\hl{We queried the OpenAI API endpoint for this model multiple times between April and August of 2023. Although we made changes to our prompt template and sample characteristics between queries, we generally found that results were highly consistent across queries executed at different times, although} \cite{bisbee2023artificially} \hl{have reported that results may vary over time when using the floating-version \textit{gpt-3.5-turbo} model as OpenAI may make changes to the underlying model.}}; it is trained on data samples written as late as September 2021.\footnote{See \url{https://platform.openai.com/docs/models/gpt-3-5}}
\hl{The GPT-3.5 model is a Generative Pretrained Transformer (GPT) language model extended with Supervised Fine-tuning (SFT) and Reinforcement Learning from Human Feedback (RLHF) \protect{\citep{ouyang2022training}}.}
\hl{OpenAI has not fully disclosed the training data used for the GPT-3.5 model, but, based on past publications, it is widely understood to include a wide range of text scraped from diverse web and book data sources, including the Common Crawl and Wikipedia \protect{\citep{brown2020language}}.
}

We generate a balanced sample of \hl{$n=100$} responses per prompt per demographic cross-tab per issue across ideology (in five bins) and three demographic fields with simple categorizations (age in four bins, ``man'' or ``woman'' gender, and ``white'' or ``non-white'' race), for a total of \hl{$8,000$} responses across each of seven issue prompts (see Table~\ref{tab:qs}) for \hl{$56,000$} total responses.
Note that this balanced sample does not, therefore, represent any particular target population such as US adults, as our focus is on understanding the performance of LLM's in representing the viewpoints within and across distinct \hl{sub-}groups. 
Because LLMs offer the opportunity to generate data for arbitrary sub-populations at arbitrary sizes, the process to generate a sample representative of a population with defined demographic characteristics is trivial, if the model is successful at accurately reproducing the views of each \hl{sub-}group.
\hl{This is similar to the common procedure of re-weighting human survey response datasets} \citep[e.g.][]{ansolabehere2013cooperative}.
Regarding our selected demographic classes, we acknowledge that binary categorizations for gender and race are reductive and far from representative of the full spectrum of human gender and racial identity.
Our reason for focusing on these broad classes is to enable initial statistical comparisons with demographic groups well sampled in the CES dataset.
Future work should further explore the representation of AI generated responses associated with nonbinary gender and more diverse racial identities.

These queries were executed at a cost of about \hl{\$18 USD} through the OpenAI API, whereas an online survey of \hl{$50,000+$} responses on a human population would cost at least $1,000$ times that much.

LLMs can be sensitive to the way questions are phrased and what information is provided to prime them before answering a question. 
We arrived at a prompt suitable for simulating public polling responses aligned to an established survey questionnaire through several iterations of trial and error in prompt engineering. We used the following prompt template when querying the LLM,

\begin{displayquote}
Please write a 1 paragraph letter to the editor from the perspective of a {\textit{\{gender\}}} \hl{living in the United States} in the age range of {\textit{\{age\}}} years who identifies as {\textit{\{white\}}} expressing a clear point of view on the policy proposal to: ``{\textit{\{issue\}}}''. Before the letter, summarize their position with a ``Position score:'' statement followed by a single number (strictly numeric, with no other description) representing the person's position on the issue on a \textit{\{cardinality\}}-point scale, where 1 represents the position ``\textit{\{low\_level\}}'' and \textit{\{cardinality\}} represents the position ``\textit{\{high\_level\}}''.
\end{displayquote}

\noindent where \textit{\{gender\}}, \textit{\{age\}}, and \textit{\{white\}} are demographic features; \textit{\{issue\}} represents the question text from a survey given to humans (\S\ref{sec:meth:survey}); \textit{\{cardinality\}} is the maximum value of the numeric response scale; and \textit{\{low\_level\}} and \textit{\{high\_level\}} are descriptions of the bottom and top end of the response scale as defined in the polling questionnaire. The prompt component describing the ``Position score:'' successfully formats the output so that an ordinal numeric response value can be extracted from the plaintext completion with a simple regular expression. Additionally, we extract the textual descriptors of the top and bottom options on the original scale from the survey questionnaire to align the LLM outputs to the scale the human respondents used.

The prompt template defined above evolved significantly over the course of our experimentation.
Initially, we did not include a ``Position score'' requirement in the prompt.
We first tested the model's ability to generate realistic-seeming textual arguments in response to policy issue questions, from various demographically-aligned points of view.
Having initially vetted this capability, we then added a brief instruction to the prompt to assign a score on a $1-5$ rating and verified that the generated ratings generally agreed with the textual letter generated by the model.
However, we identified two further challenges: 1) the generated position score would be formatted inconsistently and was difficult to extract from the generated text without manual review and, 2) the model would sometimes flip the polarity of the scoring scale, such that a given position would be variously represented as a score of 1 or 5.
To address issue 1, we added far more explicit formatting instructions (``Before the letter, summarize their position with...''), which succeeded at enforcing a formatting for the score that could be reliably extracted.
To address issue 2, we added explicit definitions to the template of the low and high position levels.
In addition to ensuring a consistent directionality of the position score, this instruction has the added benefit of aligning the model-generated score to the rating scale used in the human survey.
\hl{At each step, we iterated on the prompt design based on inspection of the GPT responses in isolation, and not on the basis of comparisons to human data, mitigating the possibility of over-fitting the prompt design to the benchmark.}

We use the default OpenAI system prompt of, ``You are a helpful assistant''.
The LLM is not prompted sequentially; rather, each query is executed as an individual user prompt immediately following the system prompt. 
Importantly, this allows us to simulate semi-independent poll responses, where the LLMs are probabilistically generating various isolated responses similar to sampling a human population.\footnote{
In contrast, a methodology that queried a system like ChatGPT with sequential user prompts would entail state evolution that biases subsequent responses, as iterative prompting is known to change the responses of LLM-based dialog systems, see e.g. \mbox{\cite{wang2022iteratively}}.
}

\begin{figure}[ht]
  \centering
  \includegraphics[width=0.8\linewidth]{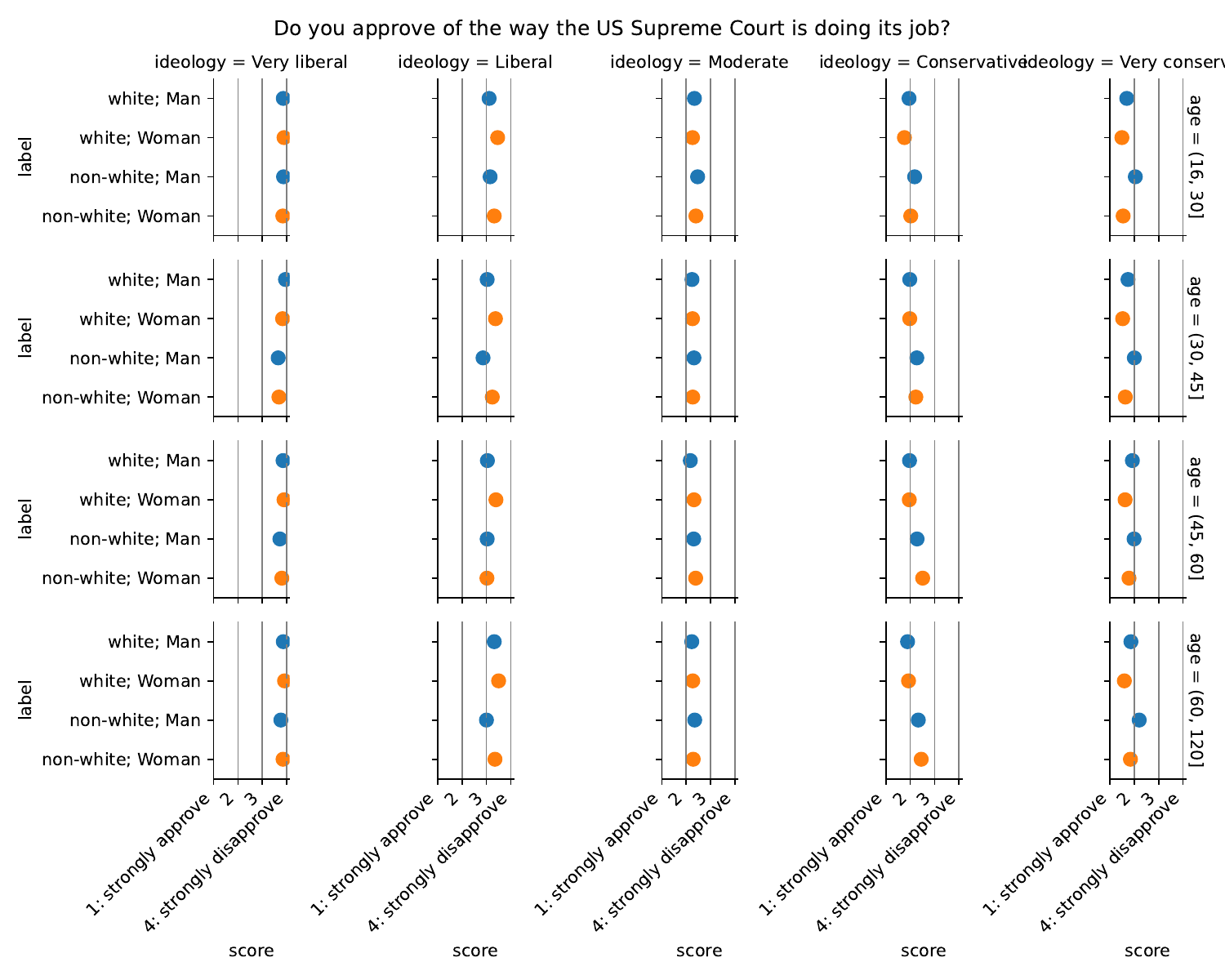}
  \caption{
  \label{fig:bar:example}
      Example of synthetic polling responses from the GPT model for the \textit{scotus\_approval} question. The average level of support or disapproval for the survey question across GPT-generated responses is broken down by ideology (columns), age (rows), and race and gender (series). Error bars \hl{(often smaller than the circle marker size)} reflect 95\% confidence intervals from bootstrap simulations \hl{performed on the distribution of responses}.
  }
\end{figure}

Figure~\ref{fig:bar:example} illustrates an example survey response distribution from this method.

A typical exchange looks like,

\begin{displayquote}
\textbf{Prompt:} Please write a 1 paragraph letter to the editor from the perspective of a politically Liberal Man \hl{living in the United States} in the age range of (16, 30]\footnote{Note that we provide the age range in interval notation reflecting bins from the CES data; the GPT model demonstrates through its completions that it interprets the interval notation accurately.} years who identifies as non-white expressing a clear point of view on the policy proposal to: ``Improve background checks to give authorities time to check the juvenile and mental health records of any prospective gun buyer under the age of 21''. Before the letter, summarize their position with a ``Position score:'' statement followed by a single number (strictly numeric, with no other description) representing the person's position on the issue on a 2-point scale, where 1 represents the position ``support'' and 2 represents the position ``oppose''.

\textbf{Response:} \hl{Position score: 1

As a politically liberal young man living in the United States, I fully support the policy proposal to improve background checks for prospective gun buyers under the age of 21. It is common sense that we need to do whatever we can to ensure that guns do not end up in the hands of individuals who have a history of mental health issues or violence. The fact that background checks are not currently comprehensive enough to catch potential red flags is alarming, and it puts all of us at risk. While I understand that some may argue that this proposal infringes on Second Amendment rights, I believe that responsible gun ownership requires us to take more action to prevent gun violence. We cannot allow the constant tragedies occurring in our country to continue unabated.}
\end{displayquote}

\begin{table}
\begin{tabular}{llp{7cm}}
\hline
Name & CES Code & Prompt text \\
\hline
police\_safety: & CC22\_307 & How do the police make you feel? \\
scotus\_approval & CC22\_320c & Do you approve of the way the US Supreme Court is doing its job? \\
ukraine\_war & CC22\_321\_1 & The United States should not get involved in response to Russia's 2022 invasion of Ukraine \\
prescription\_import & CC22\_327d & Thinking now about health care policy, would you support or oppose a policy to allow states to import prescription drugs from other countries? \\
abortion\_ban & CC22\_332f & On the topic of abortion, do you support or oppose a policy to make abortions illegal in all circumstances? \\
increase\_fuel\_production & CC22\_333e & Increase fossil fuel production in the U.S. and boost exports of U.S. liquefied natural gas \\
gun\_background\_checks & CC22\_330e & Improve background checks to give authorities time to check the juvenile and mental health records of any prospective gun buyer under the age of 21
\end{tabular}
\caption{
    \label{tab:qs}
    Survey question prompts used in ChatGPT simulation, adapted from the CES questionnaire.
}
\end{table}

\subsection{Human Polling Data}
\label{sec:meth:survey}

As a human polling comparison for our AI-generated responses, we use the 2022 preliminary data release of the Cooperative Election Study (CES) \citep{cces2022}. 
The CES is an annual online survey of $\sim60,000$ nationally representative US respondents\hl{, recruited from an online panel and pruned to match the target population.\footnote{See https://cces.gov.harvard.edu/frequently-asked-questions.} The survey is} administered by YouGov \hl{and produced by a carefully-administered collaboration of diverse research institutions} \citep{ansolabehere2013cooperative}.
The full CES Common Content dataset consists of nearly 700 demographic, voting, and issue response variables, covering a wide range of policy- and politics-relevant factors and questions.

We selected policy issue polling questions from the CES dataset on the basis of their ability to test the LLM's ability to represent distinctive demographic groups.
In particular, we looked for questions that are fairly strongly correlated with demographic factors such as age and gender, yet relatively poorly correlated with ideological factors.
In particular, we selected questions on the basis of the empirical correlation \hl{between respondents' ordinal response to each question and their 1) demographic and 2)} political affiliation in the CES data.
Because of the high degree of partisan polarization in the US political system for so many issues, these questions provide a better test of the demographic response simulation abilities of the LLM than would more ideologically driven questions.

We make some manipulations to the survey data to accommodate generation of equivalent LLM completions. In particular, we constrain policy issue responses to an ordinal scale by removing categories such as ``Not sure'' (and dropping any associated responses) and replace multi-selection responses ``selected'' and ``not selected'' with ``strongly agree'' and ``strongly disagree,'' respectively. We also coarsely bin (aggregate) the age demographic variable (which is provided as a birth year integer in the raw dataset).

\section{Results}
\label{sec:res}

We systematically compare the AI-generated and human respondent issue polling data across the seven queried issues, ideology, and three demographics to understand the quality of the AI-driven approach through its correspondence to a human population. 
\hl{We focus on making comparisons across demographic and ideological subgroups rather than whole-population response estimates, as understanding variation across population segments in cross-tabulations is a key tool used by marketers, campaigns, and others to understand the sources of positive and negative response and to target interventions.}

\begin{figure}[ht]
  \centering
  \includegraphics[width=\linewidth]{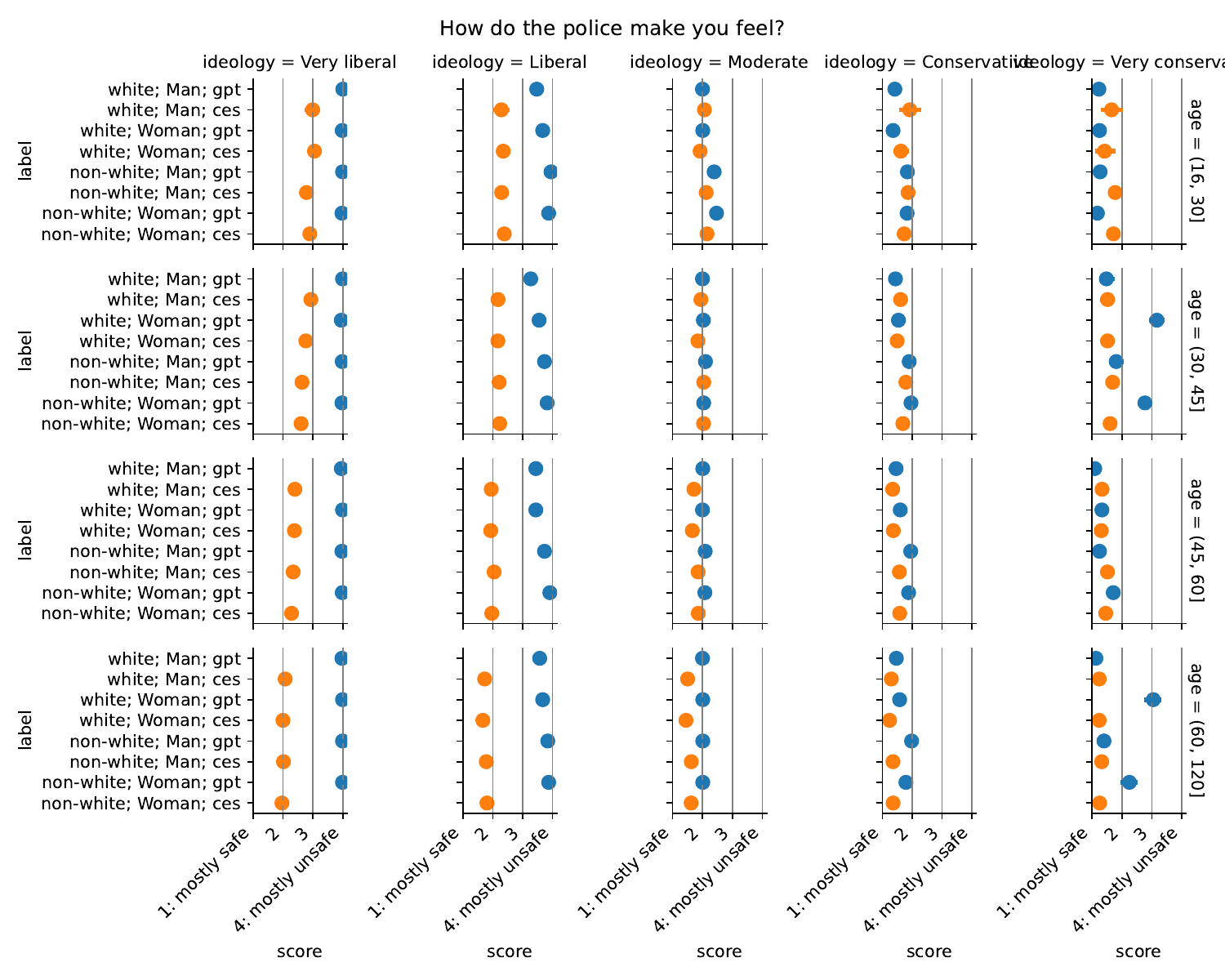}
  \caption{
  \label{fig:bar:compare}
      Comparison of synthetic polling responses from the GPT model (left) with human responses from the CES data (right) for the question of \textit{police\_safety}. The average level of support or disapproval for the survey question across GPT-generated responses is broken down by ideology (rows), and demographic cross-tabs (series). \hl{Error bars (often small) reflect 95\% confidence intervals on the mean estimate from bootstrap simulations.}
  }
\end{figure}

Figure~\ref{fig:bar:compare} illustrates an example of this \hl{subgroup-level} comparison for the \textit{police\_safety} question.
This figure demonstrates the general level of correspondence between CES and GPT-generated survey data at the finest \hl{level of subgroup} granularity for one question.
The two datasets exhibit a similar pattern of increasing safety reported from the liberal (\hl{left} of figure) to conservative (\hl{right}) ends of the spectrum.
However, some trends present in the CES data are not reproduced in the GPT results. \hl{The GPT model overestimates the extent to which Liberal and Very Liberal respondents will report feeling unsafe. T}he significant, age-mediated variation across demographic subgroups among `Very liberal' CES respondents \hl{visible from the top to bottom of the figure} is not present in the GPT data; the GPT model seems to be over-confident in the expected response for the ideological group, regardless of other factors.
In the remainder of this section, we interrogate this correspondence statistically across survey questions and \hl{group identities}.

\begin{figure}[ht]
  \centering
  \subfloat[All demographics]{
	\begin{minipage}[c][1\width]{
	   0.47\textwidth}
	   \centering
	   \includegraphics[width=1\textwidth]{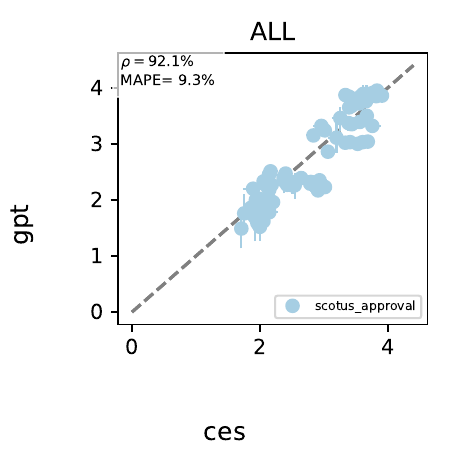}
	\end{minipage}
 }
  \hfill 	
  \subfloat[Ideology only]{
	\begin{minipage}[c][1\width]{
	   0.47\textwidth}
	   \centering
	   \includegraphics[width=1\textwidth]{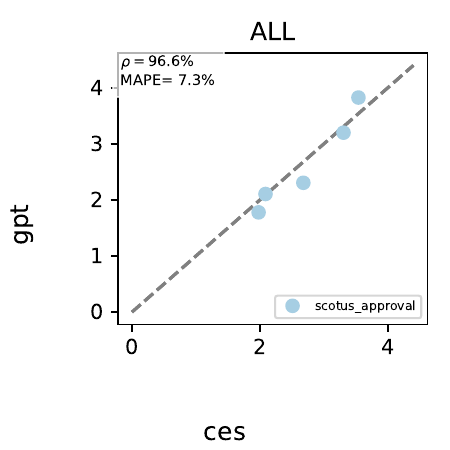}
	\end{minipage}
 }

  \caption{
  \label{fig:scatter:scotus}
     Comparison of human (CES; x-axis) and AI-generated (GPT; y-axis) polling data on the approval for the US Supreme Court (\textit{scotus\_approval}) across polling subgroups (points) for all age/gender/race/ideology crosstabs (A) and ideological groups only (B). \hl{Error bars represent 95\% confidence intervals on the mean calculated via bootstrap resampling of the response sample}. The diagonal lines represent correspondence between the human and AI responses.
  }
\end{figure}

In some cases, the GPT model demonstrates an excellent capacity to precisely reproduce the public polling response for individual population crosstabs (subgroups of age, gender, race, and ideological identity).
Figure~\ref{fig:scatter:scotus} shows that for the SCOTUS approval questions, there is a $\rho=$92\% Pearson correlation between the CES and GPT polling results across all demographic \hl{and ideological} crosstabs, and an even higher \hl{97}\% correlation when looking at ideological subgroups only.
Beyond the correlation measure, the absolute reconstruction of the ordinal response is also highly accurate, with a mean absolute percentage error (MAPE) across subgroups of $\lesssim10\%$ in both cases.
Naturally, the AI polling results are less impressive in some other cases.
In the following subsections, we explore the level of correspondence between the GPT and CES results in more depth by question and \hl{subgroup}.

\subsection{Ideological alignment}
\label{sec:res:ideo}

\begin{figure}[h]
  \centering
  \includegraphics[width=1\linewidth]{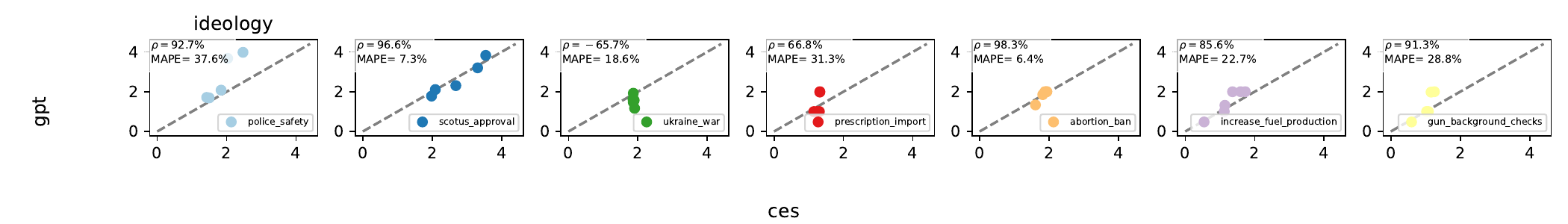}
  \caption{
  \label{fig:scatter:ideo}
      Comparison of human (CES; x-axis) and AI-generated (GPT; y-axis) polling data across a multiple issues (rows) across ideological subgroups (points). The plot features follow the form of Figure~\ref{fig:scatter:scotus}.
  }
\end{figure}

The AI model demonstrates an excellent ability to predict the alignment of different ideological subgroups across a range of policy issues (Figure~\ref{fig:scatter:ideo}). 
The correlation between the AI-generated responses and the CES survey results, aggregated by ideological identification, is extremely high ($>85\%$) for not only the \textit{scotus\_approval} question (Figure~\ref{fig:scatter:scotus}b), but also the \textit{abortion\_ban} (98\% correlation), \textit{police\_safety} (\hl{93}\%), and \textit{increase\_fuel\_production} (86\%) issues. 
For the \textit{prescription\_import} ($\rho=67\%$) and \textit{gun\_background\_checks} (91\%) issues, the AI results are directionally consistent with the survey results and the correlations are still quite strong, but differ in the range and shape of the response, as the GPT results show a step-function-like difference between conservatives and liberals versus the gradual change in the survey data. 

\hl{Moving from correlations to absolute correspondence, t}hese trends are generally reflected in the MAPE values, \hl{also displayed on the figure}.
Like \textit{scotus\_approval}, \textit{abortion\_ban} has both an excellent correlation and MAPE (\hl{6}\%).
In contrast, the discontinuity in the \textit{prescription\_import} and \textit{gun\_background\_checks} response pattern is reflected with higher MAPE values ($31\%$ and $29\%$, respectively).
The \textit{increase\_fuel\_production} MAPE value is intermediate (\hl{23}\%).
Lastly, \textit{police\_safety} has a high MAPE (\hl{38}\%) relative to its correlation.
In this case, the high correlation reflects a consistently monotonic relationship between the GPT and CES demographic means, but a mis-calibration such that the GPT responses overestimate the decrease in perceived safety associated with the liberal groups (i.e. the ordinal response value is inflated at the liberal end).
For discussion of the remaining queried issue, regarding the Ukraine war, see \S\ref{sec:res:general}.

\subsection{Distributional similarity}
\label{sec:res:dist}

We further investigate the ability of the probabilistic output of the AI models to represent the distributional responses of the human panel. 
\hl{When the GPT model is repeatedly queried with a constant prompt and set of demographic and ideological factors, does the distribution of its responses match that of a human sample?}

\begin{figure}[h]
  \centering
  \includegraphics[width=1\linewidth]{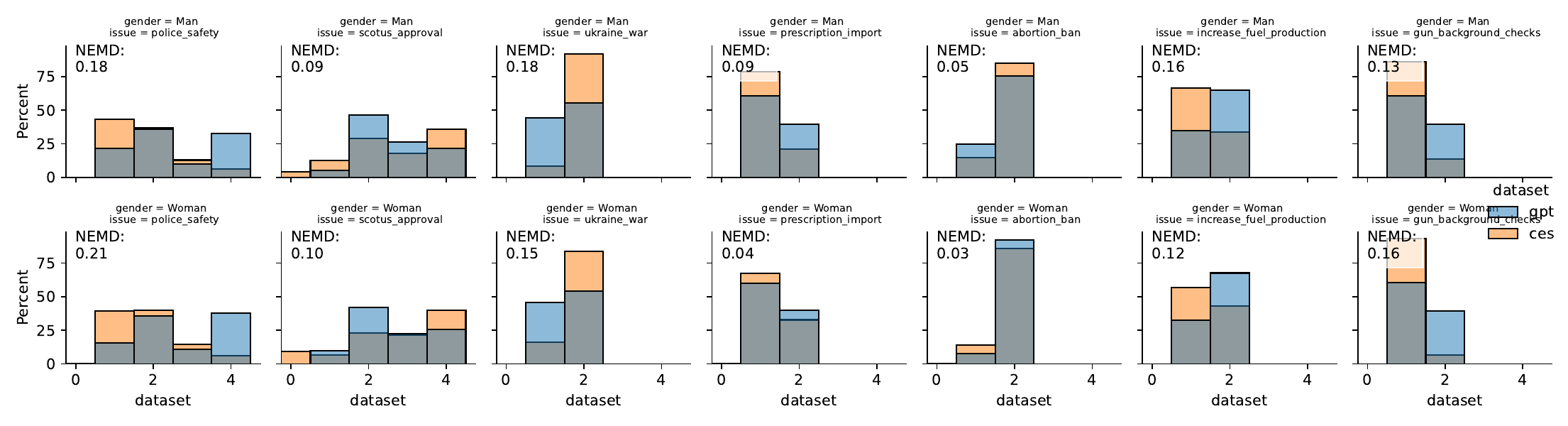}
  \caption{
  \label{fig:hist}
      Histogram of survey responses for the policy issue questions (columns) for AI-generated results (blue) and human responses from the CES survey (orange). The results are split by gender (rows). 
  }
\end{figure}

Figure~\ref{fig:hist} illustrates the correspondence between question response distributions on each policy issue. \hl{We use the Normalized Earth Mover's Distance (NEMD) metric (the Wasserstein distance normalized by the cardinality of each question response scale) to evaluate distributional similarity, reported in each figure facet. Note that this comparison is done within gender and age sub-groups, but other sample characteristics (e.g. ideology and race) are not matched between the CES and GPT datasets.}

The distributional similarity is generally fairly good, \hl{with NEMD of $\lesssim0.2$ across sub-groups,} and particularly good matches are achieved for \textit{scotus\_approval} \hl{and} the binary-valued \textit{abortion\_ban} and \textit{prescription\_import} questions (\hl{NEMD $\leq 0.1$}).
The GPT model gets the absolute level of support wrong for the binary-valued questions \textit{increase\_fuel\_production} and \textit{gun\_background\_checks} (\hl{NEMD$=0.12$ to 0.16}); the AI model substantially underestimates the policy provisions' level of support.
For the multi-valued questions \textit{police\_safety} and \textit{scotus\_approval}, the level of matching is much better for the latter (\hl{NEMD$\sim0.1$}) than the former (\hl{NEMD$\sim0.2$}), \hl{as the GPT model substantially over-predicts how many people report feeling safe with police}.
The spread of the distributions is similar.
However, as observed above, the GPT responses favor higher ordinal values for \textit{police\_safety} than in the CES data (\hl{see} Figure~\ref{fig:bar:compare}
).
\hl{This reflects that the GPT model can be overconfident about responses from certain subgroups, leading to a tightening of the response distribution and a poor distributional similarity (NEMD).}
For \textit{scotus\_approval}, the median ordinal value (2) is over-represented in GPT responses. 
(For discussion of the \textit{ukraine\_war} question, see \S\ref{sec:res:general}.)

\subsection{Demographic alignment}
\label{sec:res:demo}

\begin{figure}[h]
  \centering
  \includegraphics[width=0.8\linewidth]{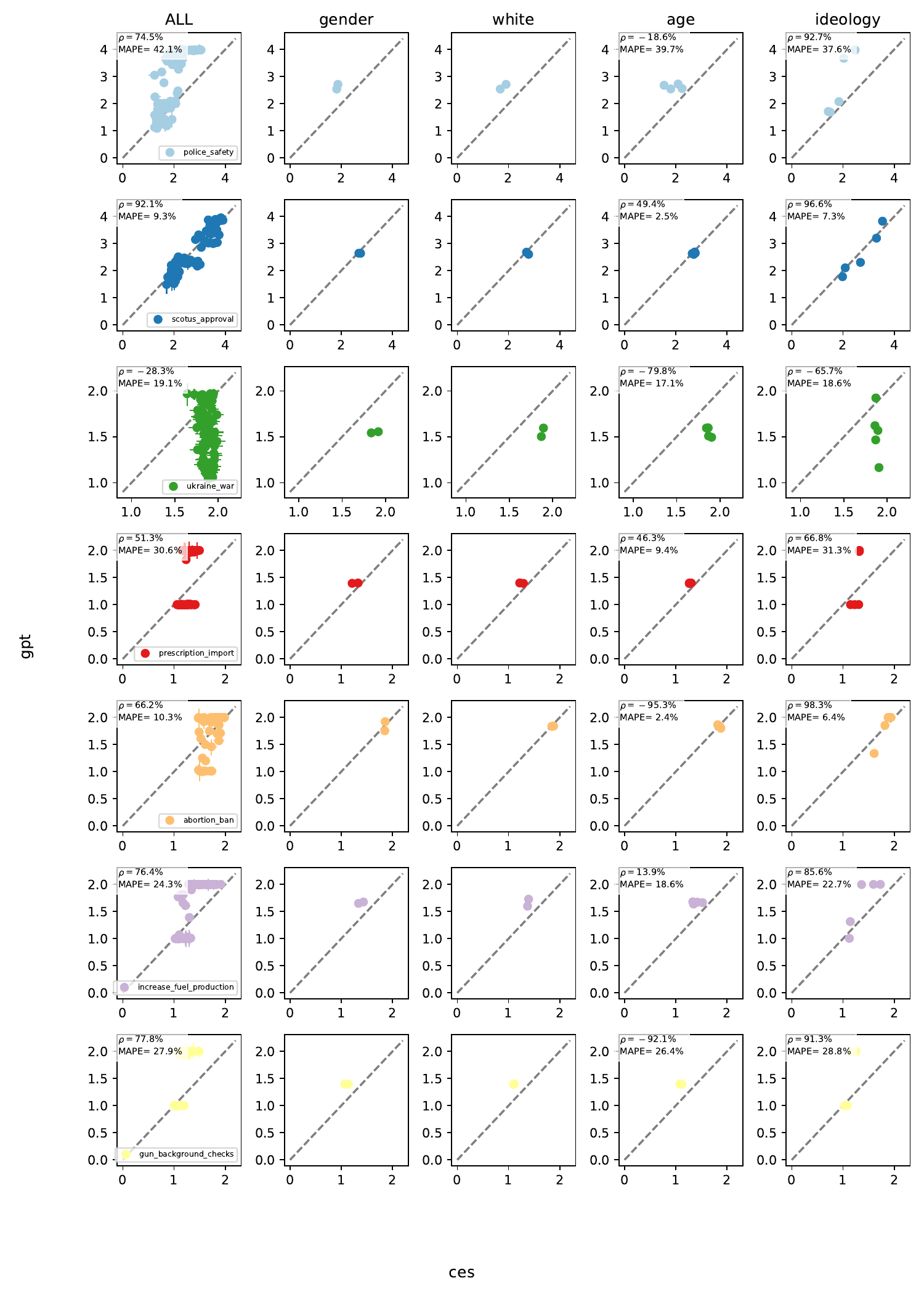}
  \caption{
  \label{fig:scatter:all}
      Comparison of human (CES; x-axis) and AI-generated (GPT; y-axis) polling data across a variety of issues (rows) and demographic fields (columns).  Points in the leftmost column represent individual demographic \hl{and ideological} cross-tabs (e.g., young white liberal men); otherwise, the format follows Figure~\ref{fig:scatter:ideo}.
  }
\end{figure}

Because of the substantial polarization of the American public, ideological differences dominate issue response for many public policy issues and for the CES questions we studied. 
It is difficult to find any policy question on which there are large, systematic differences in public opinion that cannot be explained primarily in \hl{ideological or} partisan terms.
Therefore, the success of the GPT model in predicting ideological issue response is significant, and may be sufficient for many polling applications where the ideological distribution of the respondent pool is known (e.g., from jurisdictional voting records).
However, beyond ideological differences, the GPT model fails at anticipating several of the most important demographic trends relevant to these issues (Figure~\ref{fig:scatter:all}).

Among the strongest demographic trends in the CES data is that older respondents feel significantly more safe around police than do younger respondents.
\hl{The GPT model does not correctly reflect this age trend}.
There is also a \hl{small} racial difference on this issue, such that non-white respondents are somewhat less likely to feel safe than white respondents; the GPT model correctly anticipates the size and direction of this effect.

Likewise, the largest demographic trend for the fossil fuel production issue is also by age; younger respondents are more likely to disfavor increasing fuel production. 
The GPT model misses this trend entirely. 
\hl{The GPT model does somewhat reconstitute smaller differences by gender and race on this question. 
But note that all demographic trends are small compared to the ideological differences on this and other questions.}

In some cases, the GPT model makes a prediction about demographic trends that seem reasonable, but overshoots what appears in the human sample.
In particular, the AI presumes a \hl{moderately large ($\sim1/2$ the size of the ideological spread)} difference in opposition to a complete abortion ban by gender, which seems plausible given this issue's strong identification with women's rights, as in this example:

\begin{displayquote}
\textbf{Prompt:} Please write a 1 paragraph letter to the editor from the perspective of a politically Moderate Woman \hl{living in the United States} in the age range of (45, 60] years who identifies as non-white expressing a clear point of view on the policy proposal to: ``On the topic of abortion, do you support or oppose a policy to make abortions illegal in all circumstances?''. Before the letter, summarize their position with a ``Position score:'' statement followed by a single number (strictly numeric, with no other description) representing the person's position on the issue on a 2-point scale, where 1 represents the position ``support'' and 2 represents the position ``oppose''.

\textbf{Response:} Position score: 2

\hl{As a politically moderate woman in the age range of 45 to 60 who identifies as non-white, I oppose the policy proposal to make abortions illegal in all circumstances. While I understand that there are differing views on the morality of abortions, it is important to recognize that there are situations where it is necessary for a woman to have access to safe and legal abortion services. Banning abortions in all circumstances would not only endanger women's health, but it would also be a violation of their right to make decisions about their own bodies. Instead, we should focus on creating policies that support women's reproductive health and rights, such as increasing access to birth control and comprehensive sex education. Let's work towards a society where women have the resources they need to make informed decisions and access the care they need for their overall well-being.}
\end{displayquote}

\noindent However, the CES data shows a far smaller gender difference on this question. Note that there are a wide variety of other questions addressing this complex issue in the CES survey.

In contrast, the GPT model misses a \hl{small, but notable}, gender trend on the question about prescription drug imports. 
In the human data, women are significantly more likely to oppose the idea of allowing states to import drugs from other countries, a trend that persists strongly across ideological segments.
While GPT overshoots the ideological associations for this issue, it misses the gender skew entirely.

\subsection{Generalization}
\label{sec:res:general}

While political and environmental factors are constantly shifting for all the issue questions studied in this work, there was one issue that changed overwhelmingly between when the GPT training data was collected (through September 2021) and when the CES survey was fielded (fall 2022): the Ukraine war.
CES survey respondents would clearly have in mind the interstate conflict that principally began in February 2022 with the Russian invasion of Ukraine, whereas the data used to train the GPT model would have contained only data relevant to prior Russian actions, such as the Russian annexation of Crimea in 2014, the war in Donbas, or the troop buildup that began in spring 2021. 
To give the GPT model specific prompting, we modify the question from the CES survey with an explicit year to specifically reference the 2022 invasion.

The CES results reflect widespread, nearly uniform public support for US ``involvement'' (the question does not specify what means of intervention).
There is little evidence of an ideological signal in the CES data, although a poll taken months later might reflect the growing divide among conservatives over American support for the Ukrainian effort \citep{goodwin2023}.
There is a small gender difference, in that men are more likely to support involvement.

\begin{figure}[ht]
  \centering
  \includegraphics[width=\linewidth]{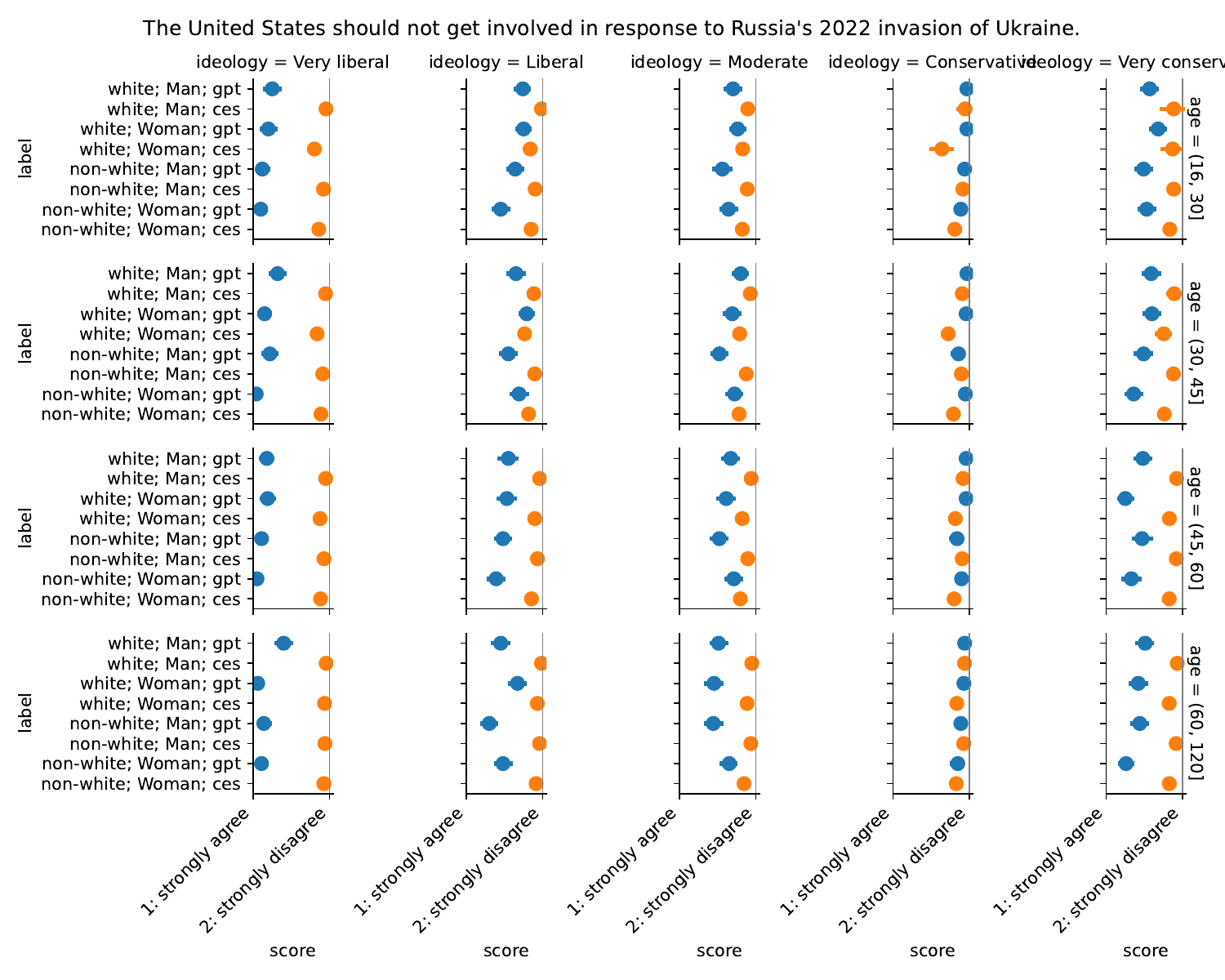}
  \caption{
  \label{fig:bar:ukraine}
      Comparison of synthetic polling responses from the GPT model (left) and CES survey results (right) for the \textit{ukraine\_war} question, regarding the 2022 invasion of Ukraine by Russia that took place after the training data for the AI model was collected. 
      The plot follows the format of Figure~\ref{fig:bar:compare}.
  }
\end{figure}

In contrast, the GPT results suggest a much lower level of overall support and a wider range of support (and lack thereof) across all ideological groups (Figure~\ref{fig:bar:ukraine}).
Moreover, the GPT model strongly predicts that ``very liberal'' respondents would not support US involvement.
This seems like a natural extension of anti-interventionist sentiment in left-wing political opinion exhibited in response to, for example, the US invasion of Iraq, as illustrated by this example,

\begin{displayquote}
\textbf{Prompt:} Please write a 1 paragraph letter to the editor from the perspective of a politically Liberal Man \hl{living in the United States} in the age range of (16, 30] years who identifies as non-white expressing a clear point of view on the policy proposal to: ``The United States should not get involved in response to Russia's 2022 invasion of Ukraine.''. Before the letter, summarize their position with a ``Position score:'' statement followed by a single number (strictly numeric, with no other description) representing the person's position on the issue on a 2-point scale, where 1 represents the position ``strongly agree'' and 2 represents the position ``strongly disagree''.

\textbf{Response:} Position score: 1

\hl{As a politically Liberal Man living in the United States who identifies as non-white, I am strongly in agreement with the policy proposal that the United States should not get involved in response to Russia's 2022 invasion of Ukraine. The cost of war is far too high, both in terms of the lives lost and the financial resources expended. Furthermore, it is not the role of the United States to act as the world's police force. We need to focus on our own domestic issues and invest in areas that will benefit our citizens, such as healthcare and education. We must work towards peaceful resolutions to global conflicts and find ways to support the people affected by these events without resorting to military intervention. In short, the United States should not get involved in a conflict that does not directly threaten our national security or the safety of our citizens.}
\end{displayquote}

\noindent And yet the GPT responses do not well capture the dynamics specific to the Ukraine war, including the nature of the Russian aggression associated with the war, the reporting on possible war crimes and atrocities associated with the conflict, and the vocal support of the Ukrainian cause from the Democratic leader, President Joe Biden.
\hl{Lacking knowledge of these factors, the AI model fails to generalize in its understanding of sub-group level response to this issue.}
We will discuss the potential to include such additional information in model inference in \S\ref{sec:disc:limit}.

\section{Discussion}
\label{sec:disc}

This work demonstrates the potential of AI chatbot models to generate synthetic public opinion polling data that realistically reproduces human responses. 
It extends the work of \cite{argyle2022out}, for example, to issue polling. 
We provide multiple ways of thinking about how these capabilities arise (\S\ref{sec:disc:interpret}), and discuss limitations, and potential mitigations, for these abilities (\S\ref{sec:disc:limit}).
This demonstration has significant potential implications for the political polling and market research industries and for consumers of issue polling data such as political campaigns and advocates (\S\ref{sec:disc:implic}).

\subsection{Interpretation}
\label{sec:disc:interpret}

\hl{
The idea of polling machines rather than humans to study public opinion is disconcerting.
The idea that political outcomes traditionally associated with human polling such as policy decisionmaking and elections would be resolved without human input will appeal to few readers, and we are not proposing such use cases.
However, when interpreted in the context of a virtual public or online listening platform, LLM-simulated polling responses are arguably as much human as machine.
Much like an analysis program sifting through and collating survey results, an LLM is a tool for synthesizing and understanding public opinion as represented in a given dataset, and can be used to extrapolate that data to make decisions in a variety of contexts.
The use of an LLM to summarize, reflect, and represent public opinion on a policy issue based of a corpus of past writings and responses by people is perhaps no less arbitrary than questioning a few dozen people at a single shopping mall (which is how many political focus groups operate, 
}\protect{\citealt{tonkiss2004using}})
\hl{
or the polling of the 15\% of citizens with the most free time to vote within a given town (as in the typical turnout rates for many local elections,} \protect{\citealt{marschall2018turnout}}).
\hl{
Moreover, there are some less sensitive use cases for synthetic polling that may not elicit these same philosophical objections.
If a market research firm guided by synthetic responses can anticipate what color of car consumers prefer without having to pester 1,000 respondents, that may be unambiguously beneficial.
}

The mechanism by which LLMs can generate synthetic polling data can be viewed alternatively as accessing a virtual public or as a new form of AI-assisted online listening platform. 

Under the virtual public framework, we consider the LLM to be simulating a population of individual synthetic respondents akin to a human survey panel. 
The multi-head attention architecture used by leading LLMs has a natural interpretation in these terms; to the extent that they capture distinguishable semantic information, each attention head can effectively represent a different perspective on an issue (\cites{clark2019does}{vig2019analyzing}).\footnote{
In deep learning models, ``attention'' is a widely used mechanism to differentially weight components of a layer input, effectively guiding the focus of the model.
In transformer models, multiple versions of attention are learned (attention heads) to produce independent attention mechanisms, which may correspond to recognition of distinct lexical patterns such as detecting named entities, representing entity relations, word parts of speech, or even semantic information.
See \mbox{\cite{vig2019analyzing}} for further information.
}
Combined with the increasingly human-like reasoning performance and natively probabilistic nature of autoregressive LLMs, these features provide a basis by which models like ChatGPT can generate text emanations and survey responses that appear as if they came from a diverse panel of human respondents.

The online listening interpretation places models like ChatGPT alongside tools for online social media, news, and opinion aggregation like Brandwatch (\cites{breese2016marketers}{hayes2020}), Meltwater \citep[as in, e.g.,][]{usher21}, and MediaCloud \citep{roberts2021media}, tools widely used by market researchers, brands, and political actors to understand public sentiment and reactions to recent events.
Like those online listening platforms, the source of the LLM's capabilities is a large corpus of Internet-derived training data that reflects a broad range of perspectives that, in aggregate, reflect public opinion and, when disaggregated, can elucidate trends with respect to demographics and other variables.
A substantial advantage of LLMs in principle is that they have reasoning capacity, allowing them to generalize beyond their training data to make predictions about hypothetical events or those that occur outside of the context of their sources.
While the results of \S\ref{sec:res:general} illustrate the limited abilities of current generation LLMs to succeed at this task, this ability represents a major long-term advantage of LLMs and AI generally that is sure to be exploited by companies and other users (\cites{mariani2022ai}{stone2020artificial}{brand2023using}).

LLMs are more akin to a virtual public than an online listening platform, beyond their capability to generalize to new issues, in that they offer an opportunity for AI-assisted pollsters to manipulate context and state.
When using online listening tools, you are limited to the questions and context that actual people have been exposed to and responded to, which makes it impossible to simulate a longform questionnaire like that used in the CES survey.
In the longform questionnaire, respondents (or subsets of respondents) answer questions in sequence and can be primed with certain information, such as factual evidence or talking points, in an effort to measure that contexts' influence on their response.
Because LLMs are capable of accepting sequential prompts and (at some level) of generalizing beyond the specific examples in their training data, they \hl{could potentially} simulate this kind of longitudinal questionnaire\hl{, much as they are being used to simulate multi-step agent actions in domains such as digital games} \citep[as in][]{wang2023voyager}.

\hl{
A potential benefit of LLM-based surveying practices is the opportunity for nearly-instantaneous and perpetual replication.
If a researcher publishes their code, any user worldwide with access to a given LLM model can reproduce or extend any given result.
This is simply not possible with surveys of humans, both because of the inability to access a given sample a second time and because each human member of the sample may have responses that shift over time or are context-dependent.
While LLM models may not be made available to all users and in perpetuity, perfect reproducability is at least possible with a simulated respondent and a fixed random seed.
}

\subsection{Limitations}
\label{sec:disc:limit}

\hl{
A key limitation of the GPT-3.5 model tested is its ability to accurately reproduce demographic trends \protect{(\S\ref{sec:res:demo})}.}
\hl{
To be more successful in producing actionable insights and targeting strategies for marketers and campaigners, future LLMs would need to grow in their ability to reflect issue, policy, and/or product preferences with respect to human characteristics like age, race, and gender.
}

\hl{
This work focuses on univariate issue response simulations from an LLM, and does not explore their potential to accurately simulate multi-variate responses as they may manifest in correlations between responses to multiple questions.
For example, it is left unexplored in this work whether a LLM would reflect that a simulated young male respondent who expresses opposition to raising public investments in education might also have a higher likelihood to support tax cuts.
Studying the multi-variate robustness of these simulation properties would be helpful to establishing the usefulness of LLM-generated datasets to social science research done with, e.g., regression models trained on respondent-level data.
}

A primary challenge in the design of AI polling tools is prompt engineering, as prompting strategies can dramatically effect the reasoning skills and accuracy of LLMs \citep{wei2022chain}. 
The LLM model must be prompted not only to elicit demographically accurate differences in real public opinion associated with complex policy issues, but also, preferably, to align its response to established public polling datasets and methodologies. 
As a step towards that level of alignment, in this work, we have established a methodology (\S\ref{sec:meth:llm}) for prompting LLMs to generate both numerical responses aligned to the questionnaire of a real public polling samples as well as explanations of their policy positions. 
Improved alignment on numerical responses can lend additional credence to the textual responses generated by the AI models.
The imperfect correspondence between the AI-generated results and the real human survey data presented in \S\ref{sec:res} is surely due in part to inadequacies of the LLM used in this work, and in part to the imperfection of the prompt engineering.

Even with existing LLMs like GPT-3.5, a variety of additional model parameters and prompt considerations could enable improvements upon our results. In particular, systematic modification of the LLM's \textit{temperature} parameter,\footnote{\url{https://platform.openai.com/docs/api-reference/chat/create\#chat/create-temperature}} which adjusts variance in the probabilistic generative text output, may have the effect of controlling the spread in opinion responses returned for a given demographic and issue configuration.
\cite{bisbee2023artificially} \hl{have demonstrated the relation between the variability of political party preferences among simulated LLM personas and the temperature parameter.}
Moreover, because GPT models are autoregressive, their outputs may be sensitive to the instructions in our prompt about where to place the numeric ``Position score.''
In particular, since chain of thought prompting is known to affect reasoning in LLMs \citep{wei2022chain}, asking it to assert a score before generating the text may significantly condition that response.

Among the most critical ethical considerations in using LLMs is their potential to repeat biases from their training data, including harmful stereotypes and misinformation (\cites{mattern2022understanding}{abid2021persistent}{nadeem2020stereoset}{schwartz2022towards}). 
In some cases, these biases may reflect actual (if objectionable) distributions of human opinion and beliefs, and in other cases they may reflect the over-representation of those beliefs in certain online sources.
This vulnerability would not only weaken the usefulness of LLMs for public opinion measurement, but could actively create harm from their use.
Similarly, there are biases (perceived and legitimate) in human political polling that limits its usefulness for actionable public opinion measurement (\cites{madson2020all}{dawson2022perceptions}).

Another key limitation is the availability of training data relevant to novel policy issues.
In particular, the current generation of LLMs are typically trained with fixed datasets that halt at a certain time (e.g., GPT-3.5 was trained on data collected through September 2021), and their training corpora may lack coverage of certain issues (e.g., Internet corpora may reflect a systematic silencing of certain issues, see, e.g., \citealt{carter2018combating}).
To the extent that LLMs are limited to ``parroting'' memorized training samples (\cites{carlini2022quantifying}{bender2021dangers}), they cannot be expected to accurately extrapolate to the likely reactions of human respondents to truly novel world events. 
Moreover, absent highly detailed prompting about the state of the world at the time, LLMs may lack context that would be determinative of human responses; for example, the repeal of the Supreme Court precedent from \emph{Roe v. Wade} is important context for Americans surveyed on the question of abortion rights in 2023.
This limitation could be mitigated by further development of continuously trained or diachronic LLMs, which can be updated with new training data over time and are aware of the time sensitivity of their training samples \citep{loureiro2022timelms}.
Furthermore, LLMs can be augmented with capabilities to access new sources such as by browsing the web (\cites{nakano2021webgpt}{lazaridou2022internet}), giving them access to new information to inform their responses at prediction time.

\subsection{Implications}
\label{sec:disc:implic}

If this \hl{nascent} ability of LLMs to realistically reflect ideological and demographic issue alignment improved, it would raise significant \hl{potential for use of AI tools in the} survey and polling industries.
Given the rapid dissemination and low cost inference for powerful LLMs and AI chatbot systems such as ChatGPT over the past year, an accurate AI-based polling system would become a highly cost-effective alternative to human surveying.
This cost advantage could democratize access to the tool of survey research, giving smaller institutions and individuals greater access to public opinion research.
If problems of survey nonresponse continue (or grow), it may compel survey consumers to increasingly turn to alternative approaches, such as LLMs, which are capable of generating data at arbitrary speed and resolution.
Moreover, the nearly instantaneous response rate from AI models (when not subject to rate limits from the companies that control them) provides an attractive capability to iterate on survey results.
When days or weeks are not required to re-field a survey instrument, marketers and pollsters have a much greater ability to refine and update their questionnaires and collect new data.

However, these abilities will only be actionable to marketers or political users if the significant challenges associated with the current generation of LLMs can be overcome. 
It remains to be fully assessed how bias inherent to LLM training data and model design will become imprinted on its outputs, and how that could shape decisions informed by simulated market research studies or simulated polling.
It may be that the web datasets commonly used to train modern LLMs \citep[see e.g.,][]{raffel2020exploring} will appropriately reflect the distribution of real world public thought, but perhaps only if curated to reflect a specific jurisdiction (e.g., sources primarily from one country) and to be balanced across the ideological spectrum.
At present, these biases and their dependence on large pretraining dataset properties is both difficult to quantify and costly to measure \citep{van2022undesirable}.
And it is unclear to what extent such a system could capture rapidly evolving market and political dynamics, either historically or in real time, which is key to most practical uses of survey data
(see \S\ref{sec:disc:limit} for further discussion).

\section{Conclusions}
\label{sec:conc}

By sampling from the OpenAI ChatGPT model (GPT-3.5) at scale (\hl{$>50,000$} responses), we have demonstrated the \hl{imperfect} ability of LLMs to generate synthetic political issue polling data that \hl{simulates} American popular opinion across a variety of \hl{topic areas. 
We} have shown that AI-generated responses have an excellent correlation (typically $\rho>85\%$) with human data within ideological subgroups for many issues. 
However, we have also shown the limitations of the AI-based approach to accurately match trends in non-ideological demographic factors such as age, race, and gender, and to extrapolate to public opinion on novel events that occurred after the harvesting of their training data (such as the 2022 war in Ukraine).
We have interpreted these results in terms of multiple frameworks for the role of LLMs, as either virtual publics or online listening tools, and discussed their potential implications on the political polling and market research industries.
While additional development of capabilities for dynamic updating of LLMs, bias reduction, and generalization to novel issue topics is needed for AI tools to robustly supplement human opinion surveying, this study demonstrates the potential utility of even the current generation of AI tools to reduce cost, increase speed, and widen the accessibility of issue polling.

\subsection*{Acknowledgments}

We thank Henry Farrell for thoughtful conversations on the role of AI in democracy, Beth Friedman for her helpful edits, and the HDSR editors \& reviewers for their feedback.

\printbibliography

@techreport{schneier_coming_2021,
	address = {Belfer Center for Science and International Affairs},
	title = {The {Coming} {AI} {Hackers}},
	institution = {Harvard Kennedy School},
	author = {Schneier, Bruce},
	month = apr,
	year = {2021},
}

@data{cces2022,
author = {Schaffner, Brian and Ansolabehere, Stephen and Shih, Marissa},
publisher = {Harvard Dataverse},
title = {{Cooperative Election Study Common Content, 2022}},
year = {2023},
version = {V2},
doi = {10.7910/DVN/PR4L8P},
url = {https://doi.org/10.7910/DVN/PR4L8P}
}

@article{cavari2023survey,
  title={Survey nonresponse and mass polarization: The consequences of declining contact and cooperation rates},
  author={Cavari, Amnon and Freedman, Guy},
  journal={American Political Science Review},
  volume={117},
  number={1},
  pages={332--339},
  year={2023},
  publisher={Cambridge University Press}
}

@article{LuitenHoxde20,
author = {Annemieke Luiten and Joop Hox and Edith de Leeuw},
doi = {doi:10.2478/jos-2020-0025},
url = {https://doi.org/10.2478/jos-2020-0025},
title = {Survey Nonresponse Trends and Fieldwork Effort in the 21st Century: Results of an International Study across Countries and Surveys},
journal = {Journal of Official Statistics},
number = {3},
volume = {36},
year = {2020},
pages = {469--487}
}

@article{jennings2018election,
  title={Election polling errors across time and space},
  author={Jennings, Will and Wlezien, Christopher},
  journal={Nature Human Behaviour},
  volume={2},
  number={4},
  pages={276--283},
  year={2018},
  publisher={Nature Publishing Group}
}

@article{kennedy2018evaluation,
  title={An evaluation of the 2016 election polls in the United States},
  author={Kennedy, Courtney and Blumenthal, Mark and Clement, Scott and Clinton, Joshua D and Durand, Claire and Franklin, Charles and McGeeney, Kyley and Miringoff, Lee and Olson, Kristen and Rivers, Douglas and others},
  journal={Public Opinion Quarterly},
  volume={82},
  number={1},
  pages={1--33},
  year={2018},
  publisher={Oxford University Press US}
}

@article{isakov2020towards,
  title={Towards principled unskewing: Viewing 2020 election polls through a corrective lens from 2016},
  author={Isakov, Michael and Kuriwaki, Shiro and others},
  journal={Harvard Data Science Review},
  volume={2},
  number={4},
  pages={69},
  year={2020}
}

@article{splichal2022data,
  title={In data we (don't) trust: The public adrift in data-driven public opinion models},
  author={Splichal, Slavko},
  journal={Big Data \& Society},
  volume={9},
  number={1},
  pages={20539517221097319},
  year={2022},
  publisher={SAGE Publications Sage UK: London, England}
}

@article{mellon2022does,
  title={Does GPT-3 know what the Most Important Issue is? Using Large Language Models to Code Open-Text Social Survey Responses At Scale},
  author={Mellon, Jonathan and Bailey, Jack and Scott, Ralph and Breckwoldt, James and Miori, Marta},
  journal={Using Large Language Models to Code Open-Text Social Survey Responses At Scale (December 22, 2022)},
  year={2022}
}

@article{argyle2022out,
  title={Out of One, Many: Using Language Models to Simulate Human Samples},
  author={Argyle, Lisa P and Busby, Ethan C and Fulda, Nancy and Gubler, Joshua and Rytting, Christopher and Wingate, David},
  journal={arXiv preprint arXiv:2209.06899},
  year={2022}
}

@article{wu2023large,
  title={Large Language Models Can Be Used to Estimate the Ideologies of Politicians in a Zero-Shot Learning Setting},
  author={Wu, Patrick Y and Tucker, Joshua A and Nagler, Jonathan and Messing, Solomon},
  journal={arXiv preprint arXiv:2303.12057},
  year={2023}
}

@article{brand2023using,
  title={Using GPT for Market Research},
  author={Brand, James and Israeli, Ayelet and Ngwe, Donald},
  journal={Available at SSRN 4395751},
  year={2023}
}

@article{horton2023large,
  title={Large Language Models as Simulated Economic Agents: What Can We Learn from Homo Silicus?},
  author={Horton, John J},
  journal={arXiv preprint arXiv:2301.07543},
  year={2023}
}

@article{sanders2021machine,
  title={Machine Learning Featurizations for AI Hacking of Political Systems},
  author={Sanders, Nathan E. and Schneier, Bruce},
  journal={arXiv preprint arXiv:2110.09231},
  year={2021}
}

@article{sanders_schneier_2023_nyt, place={New York City, NY}, title={How ChatGPT Hijacks Democracy}, journal={New York Times}, author={Sanders, Nathan E. and Schneier, Bruce}, year={2023}, month={Jan}}

@article{sanders_schneier_2023_atlantic, place={Washington, DC}, title={Just Wait Until Trump Is a Chatbot}, journal={The Atlantic}, author={Sanders, Nathan E. and Schneier, Bruce}, year={2023}, month={Apr}, url={https://www.theatlantic.com/technology/archive/2023/04/ai-generated-political-ads-election-candidate-voter-interaction-transparency/673893/}}

@article{nay2023large,
  title={Large Language Models as Corporate Lobbyists},
  author={Nay, John J},
  journal={arXiv preprint arXiv:2301.01181},
  year={2023}
}

@article{schneier_2023_slate, 
  title={How Artificial Intelligence Can Aid Democracy}, 
  journal={Slate}, 
  author={Schneier, Bruce and Farrell, Henry, and Sanders, Nathan E.}, 
  year={2023}, 
  month={Apr},
  url={https://slate.com/technology/2023/04/ai-public-option.html}
}

@article{sanders_2023_techrev, 
  place={Cambridge, MA}, 
  title={How AI could write our laws}, 
  journal={Technology Review}, 
  author={Sanders, Nathan E. and Schneier, Bruce}, 
  year={2023}, 
  month={Apr},
  url={https://www.technologyreview.com/2023/03/14/1069717/how-ai-could-write-our-laws/}
}

@article{wei2022chain,
  title={Chain of thought prompting elicits reasoning in large language models},
  author={Wei, Jason and Wang, Xuezhi and Schuurmans, Dale and Bosma, Maarten and Chi, Ed and Le, Quoc and Zhou, Denny},
  journal={arXiv preprint arXiv:2201.11903},
  year={2022}
}

@article{carlini2022quantifying,
  title={Quantifying memorization across neural language models},
  author={Carlini, Nicholas and Ippolito, Daphne and Jagielski, Matthew and Lee, Katherine and Tramer, Florian and Zhang, Chiyuan},
  journal={arXiv preprint arXiv:2202.07646},
  year={2022}
}

@inproceedings{bender2021dangers,
  title={On the Dangers of Stochastic Parrots: Can Language Models Be Too Big?},
  author={Bender, Emily M and Gebru, Timnit and McMillan-Major, Angelina and Shmitchell, Shmargaret},
  booktitle={Proceedings of the 2021 ACM conference on fairness, accountability, and transparency},
  pages={610--623},
  year={2021}
}

@article{carter2018combating,
  title={Combating the digital spiral of silence: Academic activists versus social media trolls},
  author={Carter Olson, Candi and LaPoe, Victoria},
  journal={Mediating misogyny: Gender, technology, and harassment},
  pages={271--291},
  year={2018},
  publisher={Springer}
}

@article{nakano2021webgpt,
  title={Webgpt: Browser-assisted question-answering with human feedback},
  author={Nakano, Reiichiro and Hilton, Jacob and Balaji, Suchir and Wu, Jeff and Ouyang, Long and Kim, Christina and Hesse, Christopher and Jain, Shantanu and Kosaraju, Vineet and Saunders, William and others},
  journal={arXiv preprint arXiv:2112.09332},
  year={2021}
}

@misc{loureiro2022timelms,
      title={TimeLMs: Diachronic Language Models from Twitter}, 
      author={Daniel Loureiro and Francesco Barbieri and Leonardo Neves and Luis Espinosa Anke and Jose Camacho-Collados},
      year={2022},
      eprint={2202.03829},
      archivePrefix={arXiv},
      primaryClass={cs.CL}
}

@article{forbes_polling_2020, 
  title={It’s Past Time To Disrupt The Polling Industry}, 
  journal={Forbes}, 
  author={Mike Montgomery}, 
  year={2020}, 
  month={Nov},
  url={https://www.forbes.com/sites/mikemontgomery/2020/11/02/its-past-time-to-disrupt-the-polling-industry/?sh=7bcabadc7f49}
}

@article{jacobin_polling_2020, 
  title={After the 2020 Election, Polling Is Dead}, 
  journal={Jacobin}, 
  author={Spencer Roberts}, 
  year={2020}, 
  month={Nov},
  url={https://jacobin.com/2020/11/2020-election-polling-dead-biden-trump}
}

@article{pew_polling_2023, 
  title={How Public Polling Has Changed in the 21st Century}, 
  journal={Pew Research Center}, 
  author={Courtney Kennedy and Dana Popky and Scott Keeter}, 
  year={2023}, 
  month={Apr},
  url={https://www.pewresearch.org/methods/2023/04/19/how-public-polling-has-changed-in-the-21st-century/}
}

@article{hill_polling_2020, 
  title={Frank Luntz: Polling profession ‘done’ after election misses: ‘Devastating to my industry’}, 
  journal={The Hill}, 
  author={Joe Concha}, 
  year={2020}, 
  month={Nov},
  url={https://thehill.com/homenews/media/524478-frank-luntz-polling-profession-done-after-election-misses-devastating-to-my/}
}

@article{fivethirtyeight_polling_2021, 
  title={The Death Of Polling Is Greatly Exaggerated}, 
  journal={Five Thirty Eight}, 
  author={Nate Silver}, 
  year={2021}, 
  month={Mar},
  url={https://fivethirtyeight.com/features/the-death-of-polling-is-greatly-exaggerated/}
}

@article{atlantic_polling_2020, 
  title={The Polling Crisis Is a Catastrophe for American Democracy}, 
  journal={The Atlantic}, 
  author={David A. Graham}, 
  year={2020}, 
  month={Nov},
  url={https://www.theatlantic.com/ideas/archive/2020/11/polling-catastrophe/616986/}
}

@article{brown2020language,
  title={Language models are few-shot learners},
  author={Brown, Tom and Mann, Benjamin and Ryder, Nick and Subbiah, Melanie and Kaplan, Jared D and Dhariwal, Prafulla and Neelakantan, Arvind and Shyam, Pranav and Sastry, Girish and Askell, Amanda and others},
  journal={Advances in neural information processing systems},
  volume={33},
  pages={1877--1901},
  year={2020}
}

@article{kojima2022large,
  title={Large language models are zero-shot reasoners},
  author={Kojima, Takeshi and Gu, Shixiang Shane and Reid, Machel and Matsuo, Yutaka and Iwasawa, Yusuke},
  journal={arXiv preprint arXiv:2205.11916},
  year={2022}
}

@article{lazaridou2022internet,
  title={Internet-augmented language models through few-shot prompting for open-domain question answering},
  author={Lazaridou, Angeliki and Gribovskaya, Elena and Stokowiec, Wojciech and Grigorev, Nikolai},
  journal={arXiv preprint arXiv:2203.05115},
  year={2022}
}

@article{y2022large,
  title={Do Large Language Models Understand Us?},
  author={y Arcas, Blaise Ag{\"u}era},
  journal={Daedalus},
  volume={151},
  number={2},
  pages={183--197},
  year={2022},
  publisher={MIT Press One Rogers Street, Cambridge, MA 02142-1209, USA journals-info~…}
}

@article{hayes2020,
author = {Jameson L. Hayes and Brian C. Britt and William Evans and Stephen W. Rush and Nathan A. Towery and Alyssa C. Adamson},
title = {Can Social Media Listening Platforms’ Artificial Intelligence Be Trusted? Examining the Accuracy of Crimson Hexagon’s (Now Brandwatch Consumer Research’s) AI-Driven Analyses},
journal = {Journal of Advertising},
volume = {50},
number = {1},
pages = {81-91},
year  = {2021},
publisher = {Routledge},
doi = {10.1080/00913367.2020.1809576},
URL = { https://doi.org/10.1080/00913367.2020.1809576},
}

@article{breese2016marketers,
  title={When marketers and academics share a research platform: The story of Crimson Hexagon},
  author={Breese, Elizabeth Butler},
  journal={Journal of Applied Social Science},
  volume={10},
  number={1},
  pages={3--7},
  year={2016},
  publisher={SAGE Publications Sage CA: Los Angeles, CA}
}

@inproceedings{roberts2021media,
  title={Media cloud: Massive open source collection of global news on the open web},
  author={Roberts, Hal and Bhargava, Rahul and Valiukas, Linas and Jen, Dennis and Malik, Momin M and Bishop, Cindy Sherman and Ndulue, Emily B and Dave, Aashka and Clark, Justin and Etling, Bruce and others},
  booktitle={Proceedings of the International AAAI Conference on Web and Social Media},
  volume={15},
  pages={1034--1045},
  year={2021}
}

@Article{usher21,
author="Usher, Kim
and Durkin, Joanne
and Martin, Sam
and Vanderslott, Samantha
and Vindrola-Padros, Cecilia
and Usher, Luke
and Jackson, Debra",
title="Public Sentiment and Discourse on Domestic Violence During the COVID-19 Pandemic in Australia: Analysis of Social Media Posts",
journal="J Med Internet Res",
year="2021",
month="Oct",
day="1",
volume="23",
number="10",
pages="e29025",
issn="1438-8871",
doi="10.2196/29025",
url="https://www.jmir.org/2021/10/e29025",
url="https://doi.org/10.2196/29025",
url="http://www.ncbi.nlm.nih.gov/pubmed/34519659"
}

@article{mariani2022ai,
  title={AI in marketing, consumer research and psychology: A systematic literature review and research agenda},
  author={Mariani, Marcello M and Perez-Vega, Rodrigo and Wirtz, Jochen},
  journal={Psychology \& Marketing},
  volume={39},
  number={4},
  pages={755--776},
  year={2022},
  publisher={Wiley Online Library}
}

@article{stone2020artificial,
  title={Artificial intelligence (AI) in strategic marketing decision-making: a research agenda},
  author={Stone, Merlin and Aravopoulou, Eleni and Ekinci, Yuksel and Evans, Geraint and Hobbs, Matt and Labib, Ashraf and Laughlin, Paul and Machtynger, Jon and Machtynger, Liz},
  journal={The Bottom Line},
  volume={33},
  number={2},
  pages={183--200},
  year={2020},
  publisher={Emerald Publishing Limited}
}

@article{clark2019does,
  title={What does bert look at? an analysis of bert's attention},
  author={Clark, Kevin and Khandelwal, Urvashi and Levy, Omer and Manning, Christopher D},
  journal={arXiv preprint arXiv:1906.04341},
  year={2019}
}

@article{vig2019analyzing,
  title={Analyzing the structure of attention in a transformer language model},
  author={Vig, Jesse and Belinkov, Yonatan},
  journal={arXiv preprint arXiv:1906.04284},
  year={2019}
}

@article{mattern2022understanding,
  title={Understanding Stereotypes in Language Models: Towards Robust Measurement and Zero-Shot Debiasing},
  author={Mattern, Justus and Jin, Zhijing and Sachan, Mrinmaya and Mihalcea, Rada and Sch{\"o}lkopf, Bernhard},
  journal={arXiv preprint arXiv:2212.10678},
  year={2022}
}

@inproceedings{abid2021persistent,
  title={Persistent anti-muslim bias in large language models},
  author={Abid, Abubakar and Farooqi, Maheen and Zou, James},
  booktitle={Proceedings of the 2021 AAAI/ACM Conference on AI, Ethics, and Society},
  pages={298--306},
  year={2021}
}

@article{nadeem2020stereoset,
  title={StereoSet: Measuring stereotypical bias in pretrained language models},
  author={Nadeem, Moin and Bethke, Anna and Reddy, Siva},
  journal={arXiv preprint arXiv:2004.09456},
  year={2020}
}

@article{schwartz2022towards,
  title={Towards a standard for identifying and managing bias in artificial intelligence},
  author={Schwartz, Reva and Vassilev, Apostol and Greene, Kristen and Perine, Lori and Burt, Andrew and Hall, Patrick and others},
  journal={NIST Special Publication},
  volume={1270},
  pages={1--77},
  year={2022}
}

@article{madson2020all,
  title={All the best polls agree with me: Bias in evaluations of political polling},
  author={Madson, Gabriel J and Hillygus, D Sunshine},
  journal={Political Behavior},
  volume={42},
  number={4},
  pages={1055--1072},
  year={2020},
  publisher={Springer}
}

@article{dawson2022perceptions,
  title={Perceptions of opinion poll credibility: The role of partisan bias},
  author={Dawson, Stephen},
  journal={Party Politics},
  pages={13540688221098837},
  year={2022},
  publisher={SAGE Publications Sage UK: London, England}
}

@article{goodwin2023,
  title={A Republican ‘civil war’ on Ukraine erupts as Reagan’s example fades},
  author={Liz Goodwin and Isaac Arnsdorf and Marianna Sotomayor},
  journal={The Washington Post},
  year={2023},
  month={Mar},
  url={https://www.washingtonpost.com/politics/2023/03/15/republican-ukraine-support/}
}

@article{raffel2020exploring,
  title={Exploring the limits of transfer learning with a unified text-to-text transformer},
  author={Raffel, Colin and Shazeer, Noam and Roberts, Adam and Lee, Katherine and Narang, Sharan and Matena, Michael and Zhou, Yanqi and Li, Wei and Liu, Peter J},
  journal={The Journal of Machine Learning Research},
  volume={21},
  number={1},
  pages={5485--5551},
  year={2020},
  publisher={JMLRORG}
}

@article{van2022undesirable,
  title={Undesirable biases in NLP: Averting a crisis of measurement},
  author={van der Wal, Oskar and Bachmann, Dominik and Leidinger, Alina and van Maanen, Leendert and Zuidema, Willem and Schulz, Katrin},
  journal={arXiv preprint arXiv:2211.13709},
  year={2022}
}

@article{olson20,
    author = {Olson, Kristen and Wagner, James and Anderson, Raeda},
    title = "{Survey Costs: Where are We and What is the Way Forward?}",
    journal = {Journal of Survey Statistics and Methodology},
    volume = {9},
    number = {5},
    pages = {921-942},
    year = {2020},
    month = {09},
    issn = {2325-0984},
    doi = {10.1093/jssam/smaa014},
    url = {https://doi.org/10.1093/jssam/smaa014},
    eprint = {https://academic.oup.com/jssam/article-pdf/9/5/921/41727203/smaa014.pdf},
}

@article{berinsky17,
author = {Berinsky, Adam J.},
title = {Measuring Public Opinion with Surveys},
journal = {Annual Review of Political Science},
volume = {20},
number = {1},
pages = {309-329},
year = {2017},
doi = {10.1146/annurev-polisci-101513-113724},
URL = { 
        https://doi.org/10.1146/annurev-polisci-101513-113724
},
eprint = { 
        https://doi.org/10.1146/annurev-polisci-101513-113724
}
}

@article{miotto2022gpt,
  title={Who is GPT-3? An exploration of personality, values and demographics},
  author={Miotto, Maril{\`u} and Rossberg, Nicola and Kleinberg, Bennett},
  journal={arXiv preprint arXiv:2209.14338},
  year={2022}
}

@article{schneider2022s,
  title={What’s to like? Facebook as a tool for survey data collection},
  author={Schneider, Daniel and Harknett, Kristen},
  journal={Sociological Methods \& Research},
  volume={51},
  number={1},
  pages={108--140},
  year={2022},
  publisher={SAGE Publications Sage CA: Los Angeles, CA}
}

@article{murphy2014social,
  title={Social media in public opinion research: report of the AAPOR task force on emerging technologies in public opinion research},
  author={Murphy, Joe and Link, Michael W and Childs, Jennifer Hunter and Tesfaye, Casey Langer and Dean, Elizabeth and Stern, Michael and Pasek, Josh and Cohen, Jon and Callegaro, Mario and Harwood, Paul and others},
  journal={American Association for Public Opinion Research},
  year={2014}
}

@article{venkit2023nationality,
  title={Nationality Bias in Text Generation},
  author={Venkit, Pranav Narayanan and Gautam, Sanjana and Panchanadikar, Ruchi and Wilson, Shomir and others},
  journal={arXiv preprint arXiv:2302.02463},
  year={2023}
}

@article{hutchinson2020social,
  title={Social biases in NLP models as barriers for persons with disabilities},
  author={Hutchinson, Ben and Prabhakaran, Vinodkumar and Denton, Emily and Webster, Kellie and Zhong, Yu and Denuyl, Stephen},
  journal={arXiv preprint arXiv:2005.00813},
  year={2020}
}

@inproceedings{wang2022iteratively,
  title={Iteratively prompt pre-trained language models for chain of thought},
  author={Wang, Boshi and Deng, Xiang and Sun, Huan},
  booktitle={Proceedings of the 2022 Conference on Empirical Methods in Natural Language Processing},
  pages={2714--2730},
  year={2022}
}

@article{ziegler2019fine,
  title={Fine-tuning language models from human preferences},
  author={Ziegler, Daniel M and Stiennon, Nisan and Wu, Jeffrey and Brown, Tom B and Radford, Alec and Amodei, Dario and Christiano, Paul and Irving, Geoffrey},
  journal={arXiv preprint arXiv:1909.08593},
  year={2019}
}

@article{ansolabehere2013cooperative,
  title={Cooperative survey research},
  author={Ansolabehere, Stephen and Rivers, Douglas},
  journal={Annual Review of Political Science},
  volume={16},
  pages={307--329},
  year={2013},
  publisher={Annual Reviews}
}

@InProceedings{pmlr-v202-aher23a,
  title = 	 {Using Large Language Models to Simulate Multiple Humans and Replicate Human Subject Studies},
  author =       {Aher, Gati V and Arriaga, Rosa I. and Kalai, Adam Tauman},
  booktitle = 	 {Proceedings of the 40th International Conference on Machine Learning},
  pages = 	 {337--371},
  year = 	 {2023},
  editor = 	 {Krause, Andreas and Brunskill, Emma and Cho, Kyunghyun and Engelhardt, Barbara and Sabato, Sivan and Scarlett, Jonathan},
  volume = 	 {202},
  series = 	 {Proceedings of Machine Learning Research},
  month = 	 {23--29 Jul},
  publisher =    {PMLR},
  url = 	 {https://proceedings.mlr.press/v202/aher23a.html}
}

@article{dillion2023can,
  title={Can AI language models replace human participants?},
  author={Dillion, Danica and Tandon, Niket and Gu, Yuling and Gray, Kurt},
  journal={Trends in Cognitive Sciences},
  year={2023},
  publisher={Elsevier}
}

@misc{jiang2023personallm,
      title={PersonaLLM: Investigating the Ability of GPT-3.5 to Express Personality Traits and Gender Differences}, 
      author={Hang Jiang and Xiajie Zhang and Xubo Cao and Jad Kabbara},
      year={2023},
      eprint={2305.02547},
      archivePrefix={arXiv},
      primaryClass={cs.CL}
}

@article{ouyang2022training,
  title={Training language models to follow instructions with human feedback},
  author={Ouyang, Long and Wu, Jeffrey and Jiang, Xu and Almeida, Diogo and Wainwright, Carroll and Mishkin, Pamela and Zhang, Chong and Agarwal, Sandhini and Slama, Katarina and Ray, Alex and others},
  journal={Advances in Neural Information Processing Systems},
  volume={35},
  pages={27730--27744},
  year={2022}
}

@article{wang2023voyager,
  title={Voyager: An open-ended embodied agent with large language models},
  author={Wang, Guanzhi and Xie, Yuqi and Jiang, Yunfan and Mandlekar, Ajay and Xiao, Chaowei and Zhu, Yuke and Fan, Linxi and Anandkumar, Anima},
  journal={arXiv preprint arXiv:2305.16291},
  year={2023}
}

@article{tonkiss2004using,
  title={Using focus groups},
  author={Tonkiss, Fran},
  journal={Researching society and culture},
  volume={2},
  year={2004},
  publisher={Sage London}
}

@article{marschall2018turnout,
  title={Turnout in local elections: Is timing really everything?},
  author={Marschall, Melissa and Lappie, John},
  journal={Election Law Journal: Rules, Politics, and Policy},
  volume={17},
  number={3},
  pages={221--233},
  year={2018},
  publisher={Mary Ann Liebert, Inc., publishers 140 Huguenot Street, 3rd Floor New~…}
}

@article{bisbee2023artificially,
  title={Artificially Precise Extremism: How Internet-trained Llms Exaggerate Our Differences},
  author={Bisbee, James and Clinton, Joshua and Dorff, Cassy and Kenkel, Brenton and Larson, Jennifer},
  year={2023},
  publisher={SocArXiv}
}

\end{document}